\documentclass [preprint] {revtex4}

\usepackage[latin1]{inputenc}
  \usepackage{graphicx}
\usepackage{amssymb}

\def\lp {\left( }
\def\rp {\right) }
\def\lb {\left[ }
\def\rb {\right] }

\def\m {\mu}
\def\lc {\left\{ }
\def\rc {\right\} }

\def\beq {\begin{equation}}
\def\eeq {\end{equation}}
\def\bea {\begin{eqnarray}}
\def\eea {\end{eqnarray}}
\def\ni {\noindent}
\def\nn {\nonumber}

\def\P {\Pi}

\def\d {\delta}
\def\D {\Delta}
\def\e {\epsilon}

\def\m {\mu}
\def\n {\nu}

\def\p {\pi}
\def\P {\Pi}
\def\r {\rho}
\def\s {\sigma}

\def\cK {{\cal{K}}}

\def\d{\delta}
\def\e{\epsilon}

\def\D {\Delta}

\def\m{\mu}
\def\n{\nu}

\def\p {\pi}
\def\r{\rho}
\def\s{\sigma}

\def\T {\Theta}

\def\P {\Pi}

\def\rm {\right|}

\newcommand{\dkpp}{D^+ \to  K^-\pi^+\pi^+}
\newcommand{\bppp}{B^+ \to  \pi^-\pi^+\pi^+}

\newcommand{\bpmpp}{B^\pm \to  \pi^\pm \pi^+ \pi^-}

\def\p {\pi}
\def\ni {\noindent}
\begin{document}
\title{Final state interaction on $\bppp$}

%%%%%%%%%%%%%%%%%%%%%%%%%%%%%%%%%%%%%%%%%%%%%%%%%%%%%%%%%%%%%%%%%%%%%%%%%
%\author{Brazilians}
\author{I. Bediaga and P. C. Magalh\~{a}es}
\affiliation{ Centro Brasileiro de Pesquisas Físicas - CBPF - Rio de Janeiro, RJ, Brasil}
\email[]{patricia@if.usp.br}

%%%%%%%%%%%%%%%%%%%%%%%%%%%%%%%%%%%%%%%%%%%%%%%%%%%%%%%%%%%%%%%%%%%%%%%%%
\date{\today }

\begin{abstract}
The large localized CP asymmetries observed by LHCb in three-body charmless decays of $B$ mesons brings new challenge for experimentalists and their traditional models to fit data. With higher statistics, rescattering and three-body effects, previously ignored on data analysis,  become more visible. 
The time is ripe to use better theoretical assumptions as an ingredient in order to improve the analysis.
In this paper  we address the issue of rescattering effects in the charmless three-body decays
of $B$ mesons. In particular, we study the  case of  the $\bppp$ decay and show that the presence of 
hadronic loops shifts the P-wave phase near threshold to below zero, and 
modify the position of the $\rho$-meson peak and it width, in the Dalitz plot. 
\end{abstract}

\pacs{...}

\maketitle

%^^^^^^^^^^^^^^^^^^^^^^^^^^^^^^^^^^^^^^^^^^^^^^^^^^^^^^^^^^^^^^^^^^^^^^^^^^^^^^^^^^^^^^^^^^^^^^^

%\section{Introduction}
 Three-body B and D-meson decays are known experimentally to be dominated by low energy resonances on $\p\p$, $K\p$ and $KK$ channels on Dalitz plot, usually analysed by means of isobar model.
Although the isobar model allow the resonances to interfere, it does not include coupled   channels and three-body effects. The simplicity of this model is not a problem for low statistics samples of charm and beauty decays,
producing many interesting results\cite{E791sigma, E791f0,E791kappa, Cleodppp}. However,  these effects  previously ignored on isobar model, may arise and become visible in the large LHCb data\cite{LHCbppp,LHCbkkk,LHCbcpv} and one will need to include them in the fit to data.
Therefore, is urgent to improve isobar model including final state interactions  (FSI)  effects where resonances also  play a role.

Besides the interest on light scalar resonances, the study of three-body decays of heavy meson is related to  the understanding of Charge Parity (CP) violation effects. 
 CP violation arises due to the
interference between at least two different amplitudes with different weak and strong phases. 
In charmless charged B decays,  LHCb observed a large CP asymmetries \cite{LHCbppp,LHCbkkk,LHCbcpv}  across the Dalitz plot, with regions having negative and positive values in the same decays.
Studies of $\bpmpp$, subject of this work, showed that CP violation is affected by  the strong coupling of the $\p\p$ and $KK$ channels on intermediate states of the decay from different final states\cite{LHCbcpv,IgTobias,IgTobias2}, and by the interference between  overlapping resonant amplitudes, like the S and P-wave interference around the $\r(770)$ and $f_0(980)$ resonances\cite{LHCbcpv,IgTobias, IgTobias2, BenoitBppp}. 
These studies strongly suggest that FSI may play an important role as a source of CP violation.

One can think of heavy meson three-body hadronic decays as a sequence of two processes. The first stage
consists of the heavy quark weak transition, along with local quark-gluon interactions, compounding a short-distance process that 
we refer to  as the weak vertex. This short-distance transition
is followed by the hadronization and the long-distance process of
FSI, where the three light mesons  interact in all possible ways before reaching 
the detector. These two processes involve different scales and in general are
treated separately with different theoretical frameworks.

\subsection*{ weak vertex}
The most common  way to treat B and D  weak decays  is using the factorization technique, as in Refs.\cite{BenoitBppp, DK0spp}, where the decay amplitude is represented by a product of  quark currents that do not interact strongly with each other. An effective Hamiltonian describes the transitions between the quark currents with the correct CKM matrix elements and  Wilson coefficients. The latter describe perturbatively the quark-gluon interaction and are the direct link to short-distance physics. The factorization assumption works better when the decaying particle is heavy\cite{Beneke}, what is not the case for the D meson.  
An alternative way to describe the weak vertex is the heavy mesons ChPT, developed independently by Burdman and Donoghue\cite{BD} and  Wise\cite{wise}. In this approach, the charm and bottom mesons are coupled to the light SU(3) ChPT by means of SU(3) operators. In the  effective Lagrangian, the quark-gluon interactions are hidden in coupling constants that need to be determined by experiments. 

 \subsection*{FSI}
It is important to emphasize that in three-body decays there are two distinct final state interactions mechanisms. The first one is related to the two-body  scattering and includes all possible interactions between two particles, 	
exchanging resonances and coupled channels. This pure two-body interaction  have been extensively studied with models based on ChPT\cite{GL, EGPR, Bernard, CGL} and dispersion relations \cite{Pelaez, Moussalam, Pelaez2}, applying  the constraints of analyticity and unitarity of the S-matrix. 
%These models are able to describe the scattering data up to $\approx 2$ GeV.  
Approaches that consider only this kind of FSI in three-body decays  are called (2+1), or quasi-two-body, because interactions involving  the third particle, usually referred to as ``bachelor", are ignored. %This is also the case of the isobar model and   build up by adding two-body resonances.  

However, rescattering between the bachelor and the other mesons
can  also happen and be relevant. In these three-body FSI, involving  hadronic loops, also known as triangle loops, there is a momentum  sharing between the three particles in the final state. 
The exhaustive consideration of those effects in the amplitude is possible only by means of three-body calculations such as the solution of Faddeev equation\cite{lc09}. 
The importance of three-body FSI in heavy-meson decays have been investigated recently by many different groups\cite{PRD84, PatWV,patmike,Ca, Bo,Me,Jap,satoshi, kubis}, where the $\dkpp$ decay was chosen as a golden channel.
This decay was study in a simple model that implemented three-body unitarity solving Faddeev and a compatible perturbative expansion of three-body effects on FSI\cite{lc09, PRD84, patmike}, which was recently improved in Ref.\cite{PatWV}.  In particular, Refs.\cite{PRD84, PatWV}   showed that hadron loops introduce new complex structures to the $\dkpp$ amplitude, modifying both the  S- and P-wave phase. That approach also succeeded to explain the observed discrepancy between the  $K^-\pi^+$ S-wave phase shift from scattering data\cite{LASS, Estab} and that extracted from $\dkpp$ decay\cite{E791kappa, FOCUS}. 

A more involved model to  $\dkpp$ decay was performed in Ref.\cite{satoshi}, where, inspired in $\p\p N$ system with the so called Z-diagrams\cite{Jap}, the FSI was considered as  successive rescatterings between the bachelor $\p$ with the $K\p$ system forming another resonance. In this approach all the interactions are mediated by resonances which include the $K\p$ coupled channels.  Recently, another study for the $\dkpp$ decay was performed \cite{kubis}  using dispersion relation theory on three-body systems by means of Khuri-Treiman approach\cite{KhuriTreiman}.  They consider the coupling between the $\dkpp$ and $D^+\to \bar{K}^0\p^0\p^+$ and the two-body $K\p$ couple channels. Although the work developed in Refs.\cite{satoshi} and \cite{kubis} is more complex than  those of \cite{PRD84,PatWV}, they all share the same features: the importance of hadronic loops as a three-body rescattering effect on the final amplitude.

In spite of the differences between the phase space and quark contents, B and D meson decays share the same final state process. This allows us to extend the techniques developed for D decays to  B charmless three-body decays. In this work, inspired on the importance of FSI to D meson decays, we study the relevance of a rescattering effects on the P-wave of $\bppp$ decay. If one consider that  $\p^+$ and $\p^-$ could scatter in the final state of this decay, we are including the presence of $\r$ resonance family, well known on the $\p\p$ P-wave scattering\cite{Hyams}.  Therefore, when fitting the data one needs to account for the overlap of resonances coming from  different sources: the one produced by the rescattering and the one produced directly by the B vertex. This interference  affects the values of the mass and width of the resulting resonances creating a practical problem to fit data using isobar model. 

  \section*{Rescattering effects on $B\to 3\p$}
\begin{figure}[h]
\begin{center}
\includegraphics[width=.4\columnwidth,angle=0]{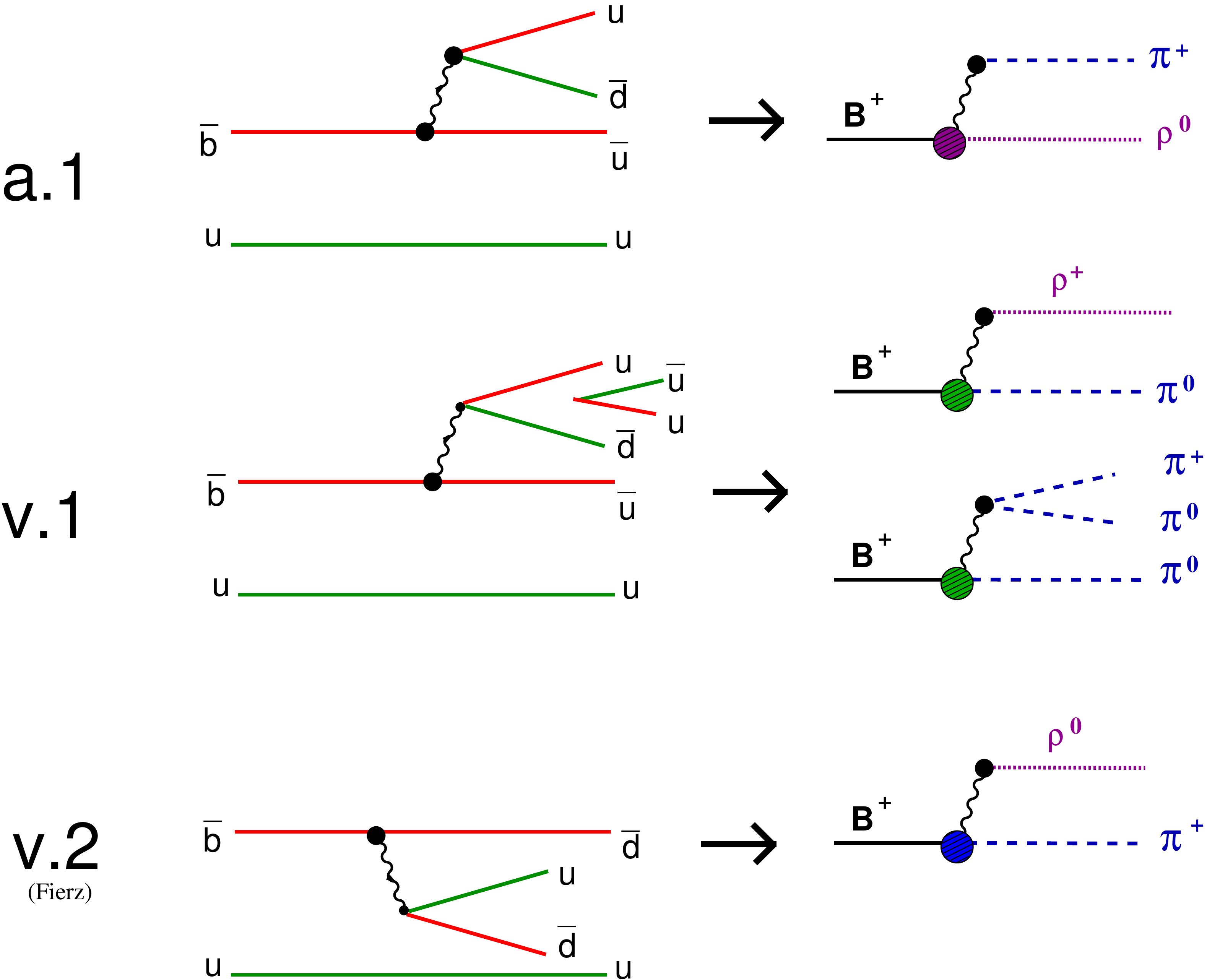}
\caption{From quarks to hadronized Feynman diagrams contribution to $\bppp$ decay. The $\r\to\p\p$ decay were omitted above.}
\label{fig:F1}
\end{center}
\end{figure}

 Chau's\cite{chau} description of heavy meson weak decays  define  the main topologies  contributing to the process $\bppp$, illustrated  in the diagrams of Fig.\ref{fig:F1}.  The diagram  $a.1$ contains an axial current, whereas  $v.1$ and $v.2$ contain vector currents. One can see that the   $v.1$ diagram does not contribute at the tree level, since it leads
 to the $\p^+\p^-\p^0$ final state, whereas in the  $v.2$ diagram, the internal $W$ emission implies  a color suppression factor.  For these reasons we choose to start our study investigating the contribution of diagram 
$a.1$.  

If one consider the evaluation of the $a.1$ tree diagram through the $\p^+\p^-$  rescattering, where by rescattering we mean the interaction between the $\p^+$ produced  from the W-boson with the $\p^-$  from the $\rho$- meson, the amplitude for $\bppp$ decay can be represented  by the series of diagrams in Fig.~2.%\ref{fig:F2}.
\begin{figure}[h]
\begin{center}
\includegraphics[width=.59\columnwidth,angle=0]{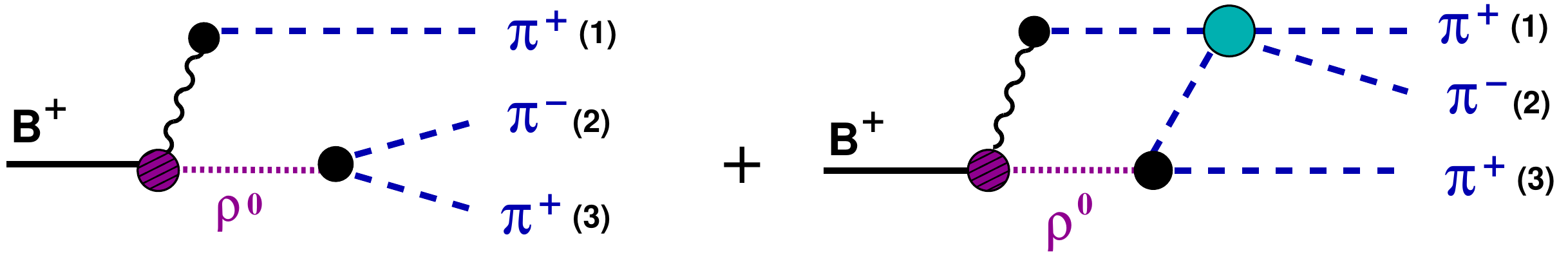}
\caption{ Diagrams contributing to $\bppp$ decay amplitude.}
\end{center}
\label{fig:F2}
\end{figure}

In this study, using a similar approach developed in Ref.\cite{PatWV}, we calculate the contribution of the diagrams in Fig.~2 to  the $\bppp$  decay amplitude. Our goal is to identify the relative importance of the rescattering effects on the whole amplitude. The contribution from the weak vertex is not included, and will be studied 
in the near future.  It is important to note that this will not affect the 
conclusions of this study, since both diagrams in Fig.~2 have the same weak vertex.

\subsection*{Tree Amplitude - $A_0$ }
The tree level contribution to the $\bppp$ decay amplitude, first diagram in Fig.~2, is given by:
\bea
iA_0 &=& i T_{B\rho}\,\,i T^\m_{W\p} \frac{i\,g_{\m\n}}{M_W^2 }\frac{i}{D_\rho} \,iT^\n_{\rho\p\p}; \
\eea
\ni where
\bea
 i T_{B\rho} &=&  -i \,\, 2m_{\rho} \,\, \frac {\epsilon^* \cdot P_B}{p_1^2}\,\, p_1\,\, F_0^{B \rho}(p_1^2)\,,
 \\[1mm]\,
 &&F_0^{B \rho}(p_1^2) =  \frac{F_{B\r}(0)}{1- M_\p ^2/m^{*2}_B}  \nn
\eea
is the amplitude transition for $B\to \r$, driven by the form factor  $ F_0^{B \rho}(p_1^2)$; and
\bea
\,\, i T^\m_{W\p} &=& if_\p p^\m_1 \,,\,\,\\[2mm]
 iT^\n_{\rho\p\p} &=& i C_W (-p_3 + p_2).\epsilon\,F^1_{\p\p}(s_{23}), %\\[2mm]
\eea
are, respectively, the pion production amplitude from the W-boson and the $\r\to \p\p$ production amplitude.
Summing over the $\rho$ polarisation yields
\bea 
&&\sum_{\lambda=0,\pm 1} \epsilon^{\lambda}\cdot(p_2-p_3) \epsilon^{\lambda^*} \cdot p_B=-p_1\cdot (p_2-p_3),\nn
\eea
 \ni and the tree amplitude is given by:
 \bea
 A_0 &=& C_0\,[-p_1\,\cdot(p_2-p_3)]\,F^1_{\p\p}(s_{23});\label{A0}\\[2mm]
 && C_0=\,2\,m_\rho\, f_\p\frac{C_W}{M^2_W}\,F_{B\r}(0),
 \eea
 where $C_0$ is a constant, and the product $-p_1\,\cdot(p_2-p_3) = (s_{13} - s_{12})$  is the angular distribution of the P-wave. Therefore, the tree amplitude, $A_{0}$, is a function of $s_{23}$, the invariant mass of the $\r$ decay channel. 

 \subsection*{Production amplitude: $\rho \to \p\p$ }
The  $\rho$ production amplitude is a crucial element in the description of $\bppp$ decay  for both tree and rescattering  amplitude. 
In a very general grounds, this process is given by the diagrams in Fig.~3, which take into account the point-like interaction and the vertex dressed by a loop of $\p\p$.
\begin{figure}[h]
\begin{center}
\includegraphics[width=.79\columnwidth,angle=0]{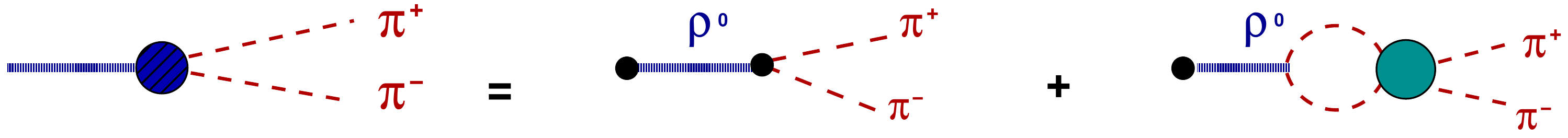}
\caption{Feynman diagrams for process $\r\to\p\p$ }
\end{center}
\label{fig:1}
\end{figure}

 The corresponding amplitude of this process is given by: 
 \beq
F^1_{\p\p}(s) = \frac{G_\r}{m_\rho^2-s} 
\lb 1 + (-\bar\Omega) T_{\p\p}\rb =
 \frac{G_\r}{m_\rho^2-s} 
\lb 1 + (-\bar\Omega) \frac{\cK}{1 + \bar\Omega\cK} \rb
= \frac{G_\r}{m_R^2-s} 
\lb \frac{1}{1 + \bar\Omega\cK} \rb .
\label{2.11}
\eeq
 \ni where $G_\r$ is the coupling $\r\to \p\p$, $\cK$ is the kernel of $\p\p$ scattering in P-wave and $\bar{\Omega}$ is $\pi\pi$ loop function.  For convenience, we re-write Eq.(\ref{2.11}) in order to include explicitly the $\p\p$ scattering phase, as done in Ref.\cite{Diogo}, and the  production amplitude becomes:
   \beq
F^1_{\p\p}(s) = g \frac{cos\d(s)}{m_\rho^2-s}\, e^{i\d(s)}
\label{2.15}
\eeq
where $\d(s)$ is the $\p\p$ phase shift. In Eq. (\ref{2.15}) we use the  CERN-Munich 
parameterization~\cite{Hyams} for the $\p\p$ phase shift,
instead of using a model to the  $\p\p$ scattering amplitude.

The production amplitude, Eq. (\ref{2.15}), is also inside the hadronic loop in  rescattering calculation. Therefore, this amplitude needs to be integrated with the loop. 
We parametrize the experimental data by a sum of poles with a complex constant in the numerator\cite{PatWV}, and write
 \bea
F^1_{\p\p}(s)= \sum_{i=1}^3 \frac{N_{\rho i}}{s - \Theta_i}\,;
\label{rhop}
\eea
where the values of $N_{\rho i} $ and $\Theta_i = a + i b$ were available in table I of Ref.\cite{PatWV}. The advantage of this form is  to allow one to use Feynman rules to integrate the loop, which will become clear in the following.

\subsection*{Rescattering Amplitude - A1}
The rescattering contribution is given by the second diagram in Fig.~2 and depends only  on the $s_{12}$, the invariant mass of the rescattered  pions. However, the $\bppp$ decay is fully symmetric and one can consider the symmetric process, given in Fig.4, in order to compare the rescattering contribution, now as a function of $s_{23}$, to the tree amplitude, calculated above. 
\begin{figure}[ht]
\begin{center}
\includegraphics[width=.35\columnwidth,angle=0]{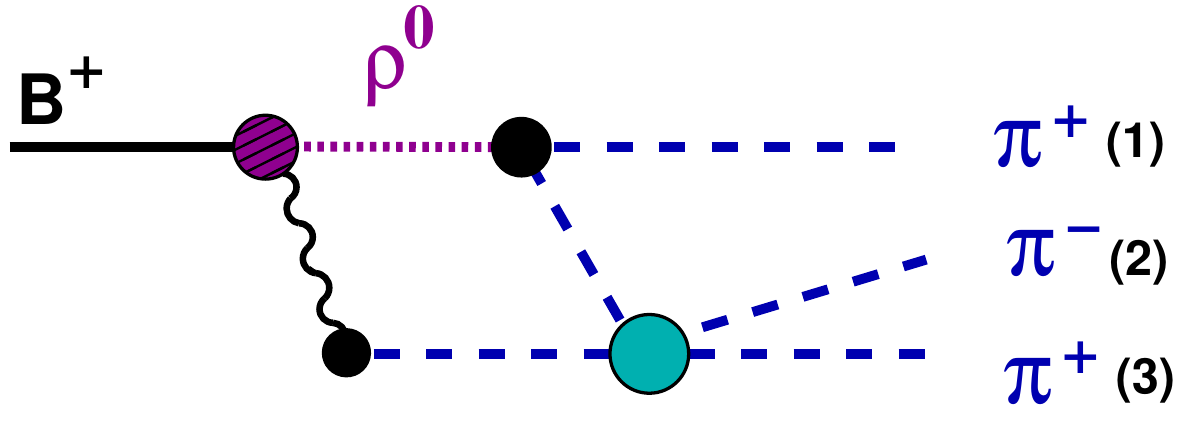}
\caption{ Rescattering diagram contribution to $\bppp$ decay. }
\end{center}
\label{fig2}
\end{figure}
Then, the rescattering contribution can be written  in terms of the tree amplitude as
\bea
A_1= -i \int \frac{d^4 \ell}{(2\p)^4} \; 
\frac{T_{\p\p} \, A_0}{\D_\p^+ \,\D_\p^- }\;,
\label{A1}
\eea
where $T_{\p\p}$ is the $\p\p$ scattering amplitude represented by the green bubble at Fig.~4 and $\D_\p^+$,  $\,\D_\p^- $ are the pions propagators inside the loop:
%\bea
$\D_\p^+ = (P_B-l)^2 - M^2_\p$ and $\D_\p^- = (l-p_3)^2 - M^2_\p$.
%\eea
It is important to note that the $B\to\r$ form factor is no longer a constant because the transferred momentum is integrated over in the loop. In this case, we use a single pole approximation and 
\bea
F_0^{B \rho}(p'^{2}_1)   = -\frac{F_{B\r}(0)\,m^{2}_{B*}}{\,\D_{B*}} ; \,\,\,\,\D_{B*} = (P_B-l)^2 -  m^{2}_{B*} \,.
\eea
The rescattering amplitude is then given by:
\bea
A_1= -i C_O (\,-m^{2}_{B*} )\int \frac{d^4 \ell}{(2\p)^4} \; 
\frac{T_{\p\p}(s_{23}) \,[-p'_1\,\cdot(p'_2-p_3)] }{\, \D_\p^+ \,\D_\p^- \,\D_{B*}}\;F^1_{\p\p}(l^2)\;,
\label{A1.1}
\eea
where  $p'_i$ are the momentum of the pions inside the loop and $F^1_{\p\p}$ is given by Eq.(\ref{rhop}). The scalar product in Eq.(\ref{A1.1}) can be written in terms of the meson propagators and the loop integral momentum $l$ as
\bea
-p'_1\,\cdot(p'_2-p_3) &=& \frac{1}{2}\,[\, \D_{\p^+} +2\,\D_{\p^-} - \,2\,s_{23} + 3\,M^2_\p + M^2_B - l^2 \,]\
\eea
and 
\bea
A_1= i \frac{\,C_O \,m^{*2}_B \,N_{\rho }}{2}\int \frac{d^4 \ell}{(2\p)^4} \; 
\lb \,T_{\p\p}(s_{23})\rb\frac{\,\lp \D_{\p^+} +2\,\D_{\p^-} - \,2\,s_{23} + 3\,M^2_\p + M^2_B - l^2 \rp}{\D_\p^+ \,\D_\p^- \,\D_{B*}\;[l^2 - \Theta_\r]} \,.\nn
\label{A1.2}
\eea
 
The $\p\p$ scattering amplitude in this channel can receive contributions from both S and P-wave, but in this study we will consider only the P-wave contribution. In this case, the s-channel amplitude projection on  P-wave result
\bea
T_{\p\p}(s_{23})\,&=&\, \frac{3\,(t -u)}{s_{23}-4\,M^2_\p\,}\,T^P_{\p\p}(s_{23});\\[2mm]
&&t-u = 2 p_1\cdot(p_2-p_3) - 2\,l\cdot(p_2-p_3) \,,
\eea
\ni where we used the CERN-Munich parametrization\cite{Hyams} to describe the  $T^P_{\p\p}$ scattering amplitude.

Finally, the rescattering amplitude for $\bppp$ is given by:
\bea
A_1 &=& i \frac{ C_O \,m^{2}_{B*}}{2\, }\,T^P_{\p\p}(s_{23}) \,N_{\rho}\,\lc\, I_1 - I^t_2 \,\, \rc,
\label{A1.3}
\eea
 where $I_1$ and $I^t_2$  are, respectively, functions of  scalar and tensorial loops integrals 
\bea
 I_1 = (s_{12} - s_{13})\,&\frac{i}{16\p^2}&\,\lc \lp  M^2_B - \,2\,s_{23} + 3\,M^2_\p + \T_i\rp \Pi_{\p^+\p^- B^*\rho_i} + \Pi_{\p^- B^*\rho_i} \right. \nn\\&&+ \left.  2\,\Pi_{\p^+ B^*\rho_i} -\,\Pi_{\p^+\p^- B*} \rc \,,
 \label{I1} \\[2mm]
 I^t_2 = 2\,(p_2-p_3)_\m\,&\frac{i}{16\p^2}&\,\lc \lp  M^2_B - \,2\,s_{23} + 3\,M^2_\p + \T_i\rp \Pi^\m_{\p^+\p^- B*\rho_i} + \Pi^\m_{\p^- B^*\rho_i} \right. \nn\\&&+\left. 2\,\Pi^\m_{\p^+ B^*\rho_i} -\,\Pi^\m_{\p^+\p^- B^*} \rc \, ,\label{I2}
\eea
\ni  and the indices on function $\Pi_{xyz}$ refer to the particles that are taking part in the loop. 
All these loop integrals can be solved using Feynman technique. A guide for 
all calculations can be found in Ref. \cite{PatThesis}.

 \section*{Results}

We compare numerically the contributions from tree, Eq.(\ref{A0}), and  rescattering amplitudes, Eq.(\ref{A1.3}),  to the total $\bppp$ decay amplitude.

\begin{figure}[h]
\hspace*{-2mm}\includegraphics[width=.5\columnwidth,angle=0]{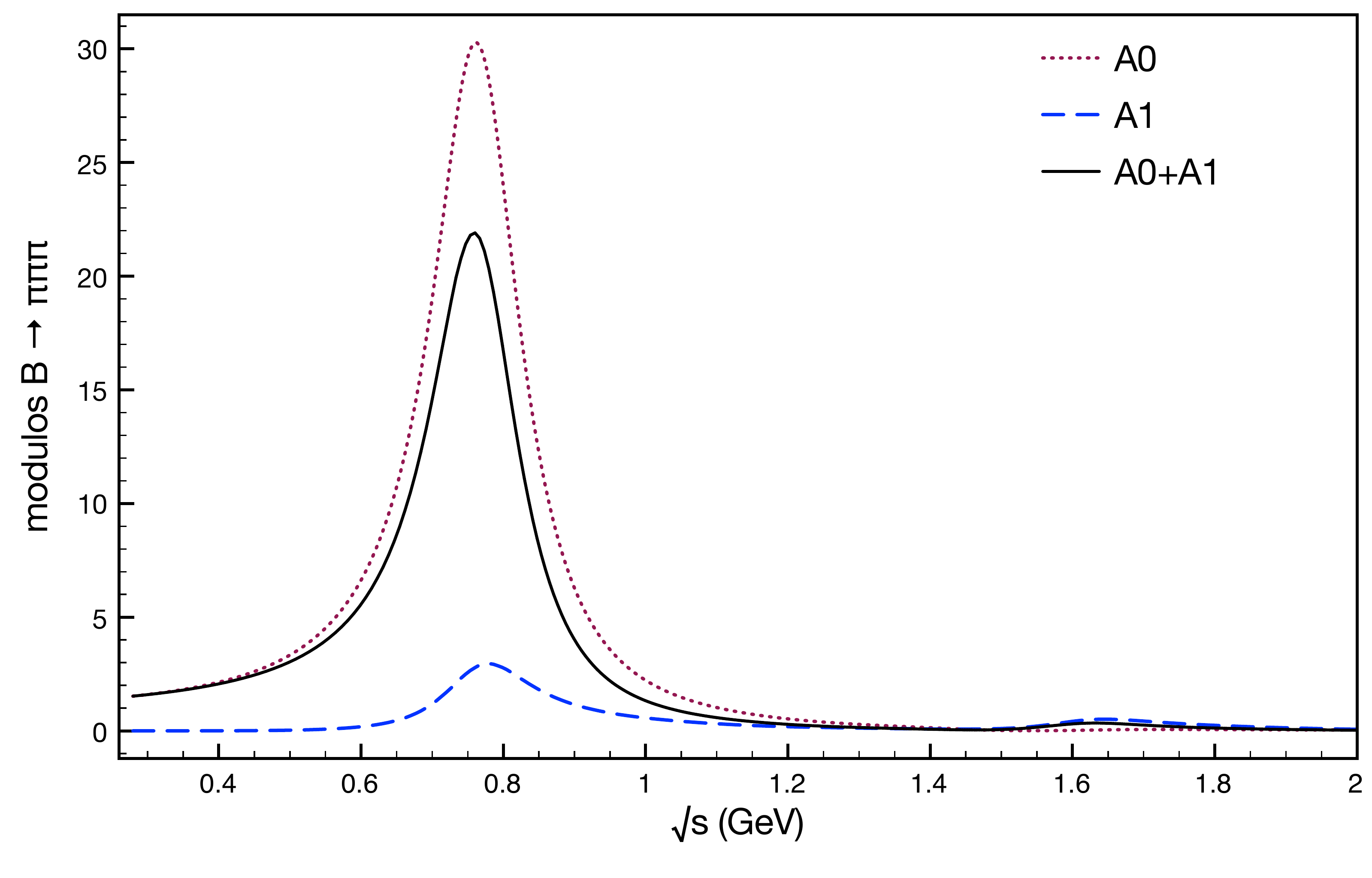}
\includegraphics[width=.5\columnwidth,angle=0]{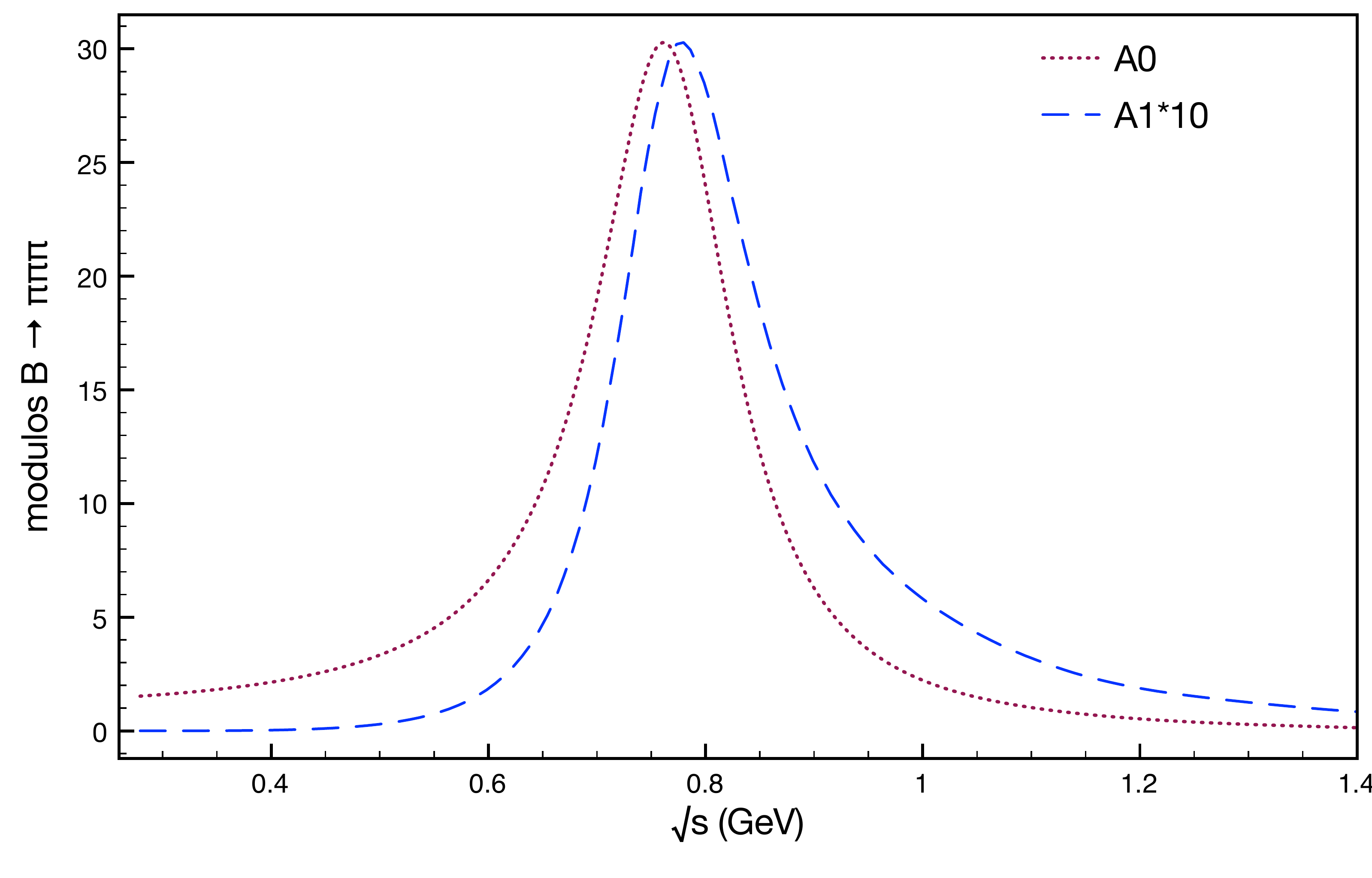}
\caption{Magnitude square of $\bppp$ decay amplitude: (left) the tree Eq.(\ref{A0}), rescattering Eq.(\ref{A1.3}) and final amplitude, (right) only tree and rescattering rescaled.}
\label{moduloZ}
\end{figure}

In Fig.\ref{moduloZ}, we show the results for the modulus what clarify the relative weight of each amplitude into the final one. In the plot on the left  one can see that the tree and rescattering amplitudes interfere destructively with the dominance of the former. Is possible to note a presence of a smaller peak around $1.7$ GeV, which is the contribution of the $\r(1450)$  coming from the $T_{\p\p}$ scattering amplitude. 
In the plot on the right, the two contributions are superimposed, with the rescattering contribution
rescaled,  showing that the line shapes are very different. Not only the position of the peak is different, 
but also the two curves are asymmetric in opposite directions, with the rescattering curve being wider than the curve from the tree amplitude.

\begin{figure}[h]
\includegraphics[width=.5\columnwidth,angle=0]{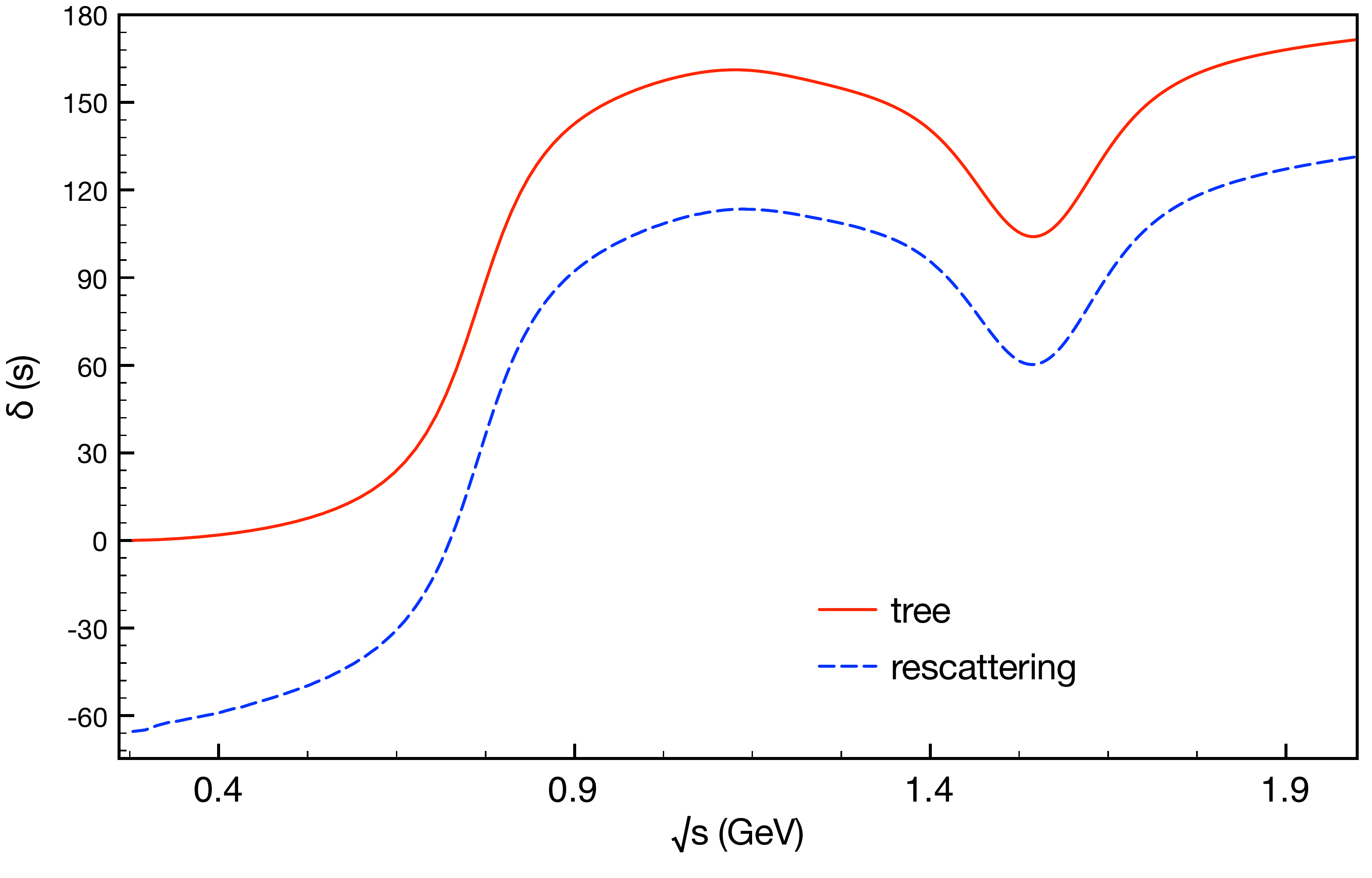}
\caption{ P-wave phase for tree, Eq.(\ref{A0}), and rescattering, Eq.(\ref{A1.3}),  contributions in $\bppp$ decay.}
\label{phase}
\end{figure}

In the Fig.\ref{phase} individual contributions to the P-wave phase are shown as a function of 
the $\p\p$ invariant mass. Although they look similar, at
threshold the rescattering contribution, Eq.(\ref{A1.3}), starts bellow $-60^o$
and crosses  $90^o$ in a different position than the tree contribution. Even if one shift up the rescattering contribution, they have a clear different energy dependence on the lower sector.
This rescattering behave at low energy is due to the hadronic loop and is in complete  agreement with what was seen in the $\dkpp$ study in Refs.\cite{PatWV,PatThesis}, the first work to show the effect of rescattering in three-body decay.

In order to analyse with more detail the effects of the rescattering when fitting the data, three samples of toy Monte Carlo with 10,000 events were generated using the software Laura++\cite{laura} to each of the amplitudes derived in the previous 
section: tree and rescattering, and to the full amplitude. In the sequence we tried to fit those amplitudes using isobar model tools, such as Relativistic Breit-Wigner\cite{BW} functions.

\begin{figure}[h]
%\hspace*{-10mm}
\includegraphics[width=.5\columnwidth,angle=0]{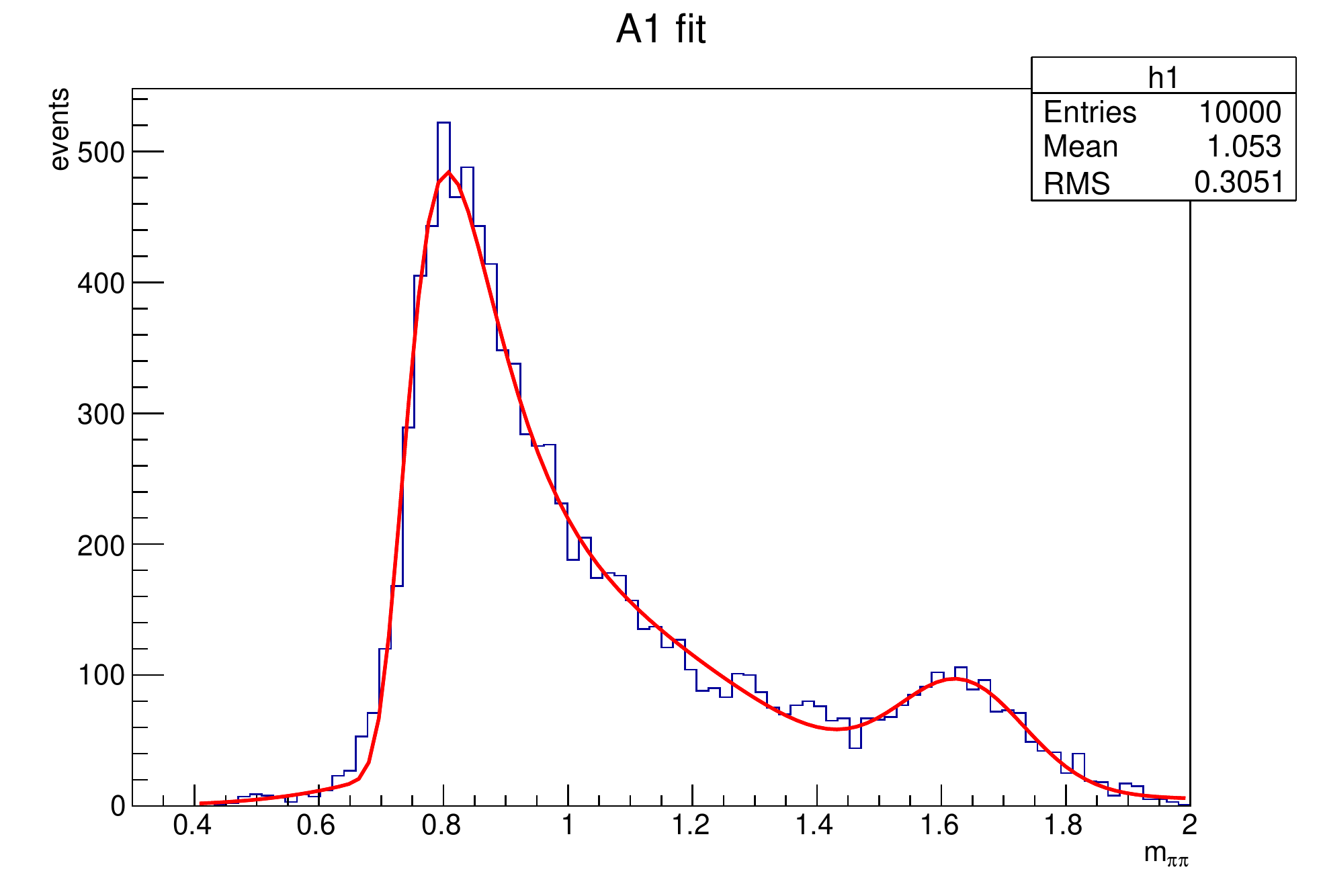}
\caption{ Result of the fit performed on rescattering contribution, Eq.(\ref{A1.3}), with two Gauss and one Landau functions.}
\label{landauGauss}
\end{figure}

The analyses on rescattering amplitude, function $A_1$, did not produced a reasonable fit with  Breit-Wigner functions.  However, we were able to parametrized this function using one Landau and two Gaussian functions, the result is shown in Fig.\ref{landauGauss}.

On the other hand, with the sample generated with only the tree amplitude, a good fit was obtained with a Relativistic Breit-Wigner\cite{BW} parametrization for the $\rho(770)$ with no barrier factors and with floating mass and width. This result is consistent once the only complex structure in the amplitude
 is the production amplitude of Eq.(\ref{2.15}). To this first fit, the results for the  resonance parameters were:  
$m_\rho = 0.775 \pm 0.001 $GeV, $\Gamma_\rho = 0.148 \pm 0.001$GeV, where the  uncertainties  are due to the sample size. Comparing to the  PDG\cite{PDG} values:  $m_\rho =0.77526 \pm 0.00025$ GeV and $\Gamma_\rho = 0.1491 \pm 0.0008$ GeV, one can see that they nicely agree.

When we add the rescattering contribution to the tree one, defining the complete amplitude to $\bppp$ decay, there is  an destructive interference around the $\rho(770)$ energy between an Breit-Wigner like function and one that is not. The effect of this interference is visible on resulting a sample that can only be fitted by a sum of two Breit-Wigner functions, with  the following parameters: 
 $m_\rho = 0.756  \pm 0.001$GeV, $\Gamma_\rho =0.163 \pm 0.003 $GeV and $m_{\rho2} =1.50 \pm 0.01  $GeV, $\Gamma_{\rho2} =0.194  \pm 0.033 $GeV; where the uncertainties  are statistical. 
The first $\rho$ contribution is clearly the result of the tree and rescattering interference, while the second $\rho$ comes only from the rescattering (Fig.\ref{moduloZ}, left). 
Comparing with the results of the tree amplitude fit, one can see that the inclusion of the rescattering 
amplitude slightly changes the $\rho(770)$ mass and increase significantly it width. The calculation of rescattering amplitude did not include any free parameter  to account for it relative weight to the tree amplitude,  what could enlarge or decrease the observed influence on changing the resonance parameters.

\section*{Final Remarks}
Our analyses on the relevance of rescattering effects in the $\bppp$ decay amplitude confirm the importance of hadronic loops as observed in Refs.\cite{PatWV, satoshi}. It shifts the phase at threshold to below zero and change the shape of the $\rho$ meson in the final amplitude.
When we include rescattering effects we allow the interference of two kind of resonances: the one produced directly from the main vertex, i.e. the one covered by  the isobar model, and the one coming from the two-body scattering, which is only possible by rescattering. 
  The above analysis show that the interference between those resonances in the same energy region change the width and the line shape of the resonance in the resulting amplitude, what become visible when performing the fit using isobar model.  This effect will also be observed on the Dalitz plot with the complexity of the other variable  interference.  In principle, if one consider the possibility of rescattering interference when performing fits to data, letting the mass and width of isobar model free, it will be possible to measure the amount of rescattering contribution in the final amplitude.
It is important  to point out that in this work we did not consider all the FSI effect nor all the dynamics in the $\bppp$. We focus on the main topologies and in the understanding of the rich FSI interference on the amplitude. 

\section*{Acknowledgements} 
We would like to thank A.C. dos Reis, A. Gomes and M.R. Robilotta for the fruitful discussions and careful reading of this manuscript. The work of PCM was supported by CNPq (Conselho Nacional de Desenvolvimento Cient\'{i}fico e Tecnol\'{o}gico). 

 \appendix
 \section{loop integrals}
For a generic triangle loop we have 
\bea
I_{xyz} &\!=\!& \frac{i}{(4\p)^2}\,\P_{x y z} \,\\[3mm] 
\P_{x y z} &\! = \!& 
 - \,\int_0^1 da\; a \, \int_0^1 db\; \frac{1}{D_{x y z}}\;,
\label{triangulo.1}\eea  
with
\bea
D_{xyz} &=& (1-a)\,m^2_x + a(1-b)\,m^2_y + ab\, m^2_z - i\e \nn \\[2mm]
&&-a (1-a)(1-b)(p_x -p_y)^2- a (1-a)b\,(p_x -p_z)^2 \nn\\[2mm]&&- a^2 b (1-b)\,(p_y -p_z)^2 ; 
\label{Dxyz}
\eea
The indices on Eqs.(\ref{I1}) and (\ref{I2}) refers to the following propagators:
\bea
\D_\p^+ &=& (P_B-l)^2 - M^2_\p;\nn\\
\D_\p^- &=& (l-p_3)^2 - M^2_\p;\nn\\
\D_{B*} &=& (P_B-l)^2 -  m^{2}_{B*} \,\nn\\
\D_\r &=& l^2 - \T_i\,; \nn
\eea
The integrals are solved numerically.

% REFERENCES OOOOOOOOOOOOOOOOOOOOOOOOOOOOOOOOOOOOOOOOOOOOOOOOOOOOOOOOOOOO

\end{document}